\documentclass[twocolumn,english,aps,graphicx,amsmath,showpacs]{revtex4}
\usepackage[T1]{fontenc}
\usepackage[latin1]{inputenc}
\usepackage{babel}
\usepackage{graphics}

\makeatletter

\providecommand{\LyX}{L\kern-.1667em\lower.25em\hbox{Y}\kern-.125emX\@}

\makeatother
\begin{document}

\title{A framework for public-channel cryptography using chaotic lasers}

\author{Noam Gross\( ^{1} \), Einat Klein\( ^{1} \), Michael Rosenbluh\( ^{1} \),
Wolfgang Kinzel\( ^{2} \), Lev Khaykovich\( ^{1} \), Ido Kanter\(
^{1} \)}

\affiliation{\( ^{1} \)Department of Physics, Bar-Ilan University,
Ramat-Gan, 52900 Israel,}

\affiliation{\( ^{2} \)Institut f\"ur Theoretische Physik,
Universit\"at W\"urzbur, Am Hubland 97074 W\"urzburg, Germany}

\begin{abstract}
Two mutually coupled chaotic diode lasers with individual external
feedback, are shown to establish chaos synchronization in the
low-frequency fluctuations regime. A third laser with identical
external feedback but coupled unidirectionally to one of the pair
does not synchronize. Both experiments and simulations reveal the
existence of a window of parameters for which synchronization by
mutual coupling is possible but synchronization by unidirectional
coupling is not. This parameter space forms the basis of a
proposed public-channel cryptographic scheme and is robust to
various possible attacks.
\end{abstract}

\maketitle

A semiconductor (diode) laser subjected to delayed optical
feedback is a well known example of a nonlinear oscillator that
displays chaotic oscillations. Different regimes of chaos,
depending on feedback strength and driving current are obtained in
lasers with external feedback. For low driving currents and
moderate feedback, characteristic intensity breakdowns known as
low-frequency fluctuations (LFFs) are typically observed
\cite{elsasser96}.

Chaos synchronization has attracted a lot of interest recently
because of its potential application in optical communication
\cite{chaos-com}. Chaos synchronization for two external-feedback
lasers coupled unidirectionally has been demonstrated
experimentally and theoretically in \cite{shore99,locquet01}.
However, mutually coupled semiconductor lasers were mainly
considered in the so called face-to-face configuration where both
lasers are solitary (i.e. they are not subjected to individual
optical or electronic feedback)
\cite{elsasser01,mutual-coupling,exception, Liu1}. These studies
revealed interesting dynamics regimes including anticipated and
retarded synchronization and leader-laggard dynamics.

One of the most studied applications of chaos synchronization in
communication systems is a private-key cryptographical system,
where the two communicating parties have a common secret key prior
to the communication process, which they use to encrypt the
transmitted messages. In optical communication, unidirectionally
coupled lasers are synchronized in a master-slave configuration,
and the secret key is the system parameters \cite{chaos-com}. The
two communicating lasers must have identical (or at least similar)
parameters, or else synchronization is impossible. The information
is added to the synchronized signal and recovered in a chaos pass
filter procedure \cite{CPF}. If an eavesdropping attacker (a third
laser), manages to reveal the parameters, the security of the
system is broken.

In many applications, the two communicating parties do not have a
common private-key, and for secure communication $must$ use
public-channel cryptography, where all the information is public.
Public-channel cryptographic methods that are used today, such as
RSA \cite{RSA}, are based on number theory, and implemented in
software. In this paper we present for the first time a framework
for all-optical $public$-channel cryptography. The advantages of
an all-optical device implemented in hardware is straightforward.

We use two mutually coupled external-feedback lasers to establish
chaos synchronization in the LFF regime. Once the two
communicating lasers are synchronized, and have identical time
varying signals, they can use the signal to mask a message in some
sophisticated way. As we are proposing a $public-channel$ system,
all the macroscopic parameters of the system are public knowledge,
and known to the attacker as well. Of course, if all the
parameters are known, the parties cannot use the signal itself to
hide the message, rather the transmitted signal consists of a
non-linear combination of time-delayed values. Such a signal
conceals the original signal, but still maintains the
synchronization.

An attacker wishing to decrypt the message must manage to
synchronize with the parties (using the same parameters), so that
he can generate the same decoding signal. We show that a third
laser with identical external feedback but coupled
unidirectionally to one of the communicating lasers, does not
synchronize with the pair. The asymmetry, whereby the two mutually
coupled lasers synchronize while the third laser does not, is
caused by the fact that the third laser is unidirectionally
coupled and does not influence the process \cite{rosen}. This
advantage of mutual coupling over unidirectional coupling appears
in a certain window of parameters.

\begin{figure}

{\centering \resizebox*{0.5\textwidth}{0.23\textheight}
{{\includegraphics{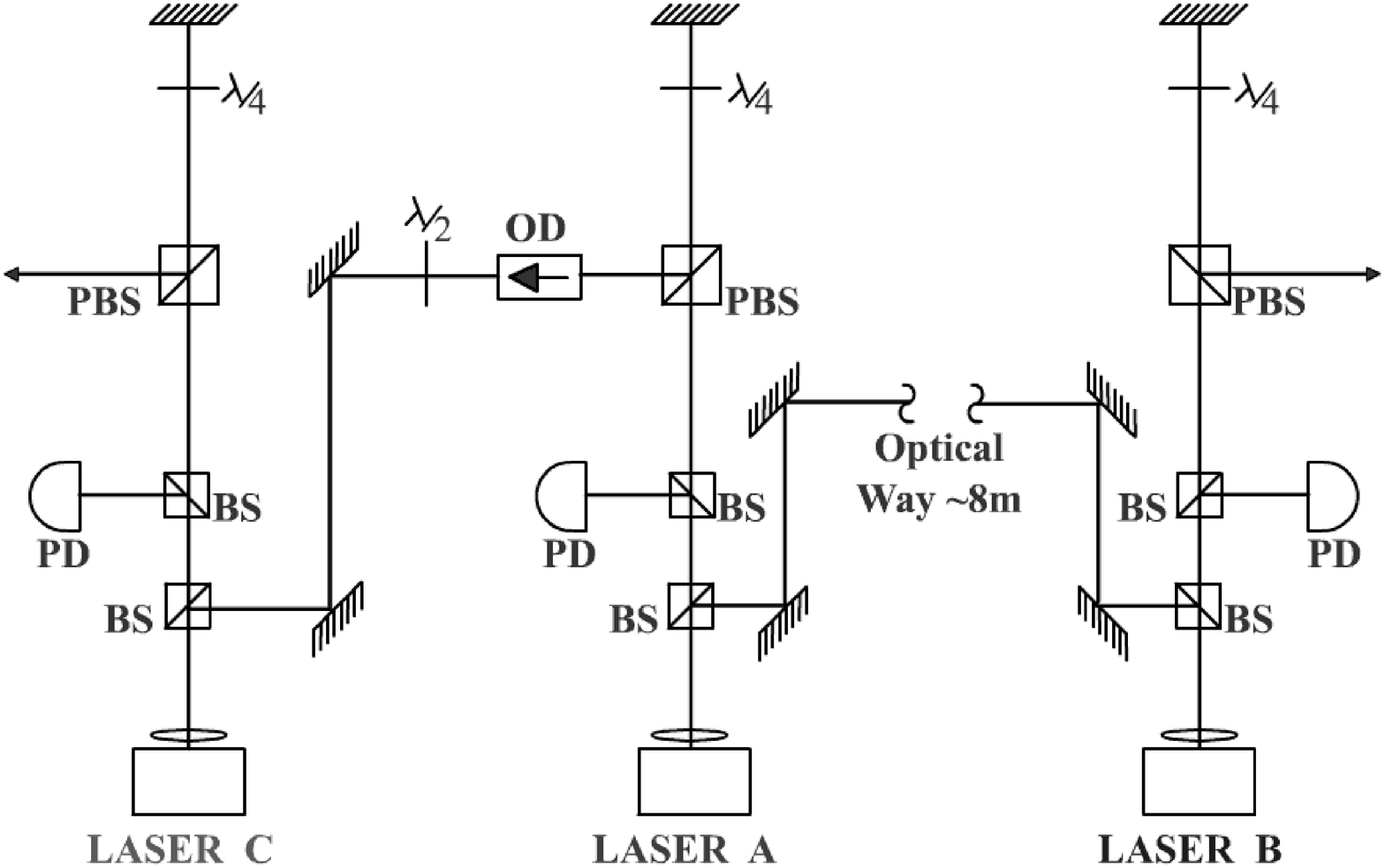}}}
\par}
\caption{\label{Laser1}Schematic diagram of the experimental
setup. Lasers A and B are mutually coupled, and C is the attacker.
BS - beam splitters; PBS - polarization beam splitters; OD -
optical isolator; PD - photodetectors. }
\end{figure}

Our experimental setup is shown schematically in Fig.
\ref{Laser1}. We use 3 single-mode lasers, $A$, $B$ and $C$,
emitting at 660 nm and operating close to their threshold. The
temperature of each laser is stabilized to better than 0.01K and
all are subjected to a similar optical feedback. The length of the
external feedback path is equal for all lasers and is set to 90 cm
(round trip time $\tau_{d}$ = 6 ns). The feedback strength of each
laser is adjusted using a $\lambda/$4 wave plate and a polarizing
beam splitter and is set to about 10$\%$ of the laser's power. Two
lasers ($A$ and $B$) are mutually coupled by injecting 5$\%$ of
the output power to each of them . The coupling optical path is
set to ~8 m ($\tau_{c}$ = 27 ns). The attacker laser ($C$) is fed
unidirectionally by one of the mutually coupled lasers.
Unidirectionality is ensured by an optical diode (-40 dB) which
prevents feedback. Three fast photodetectors (response time $<$
500 ps) monitor the laser intensities which are simultaneously
recorded by a digital oscilloscope.

\begin{figure}
\vspace{-0.5cm}
{\centering
\resizebox*{0.5\textwidth}{0.3\textheight}
{{\includegraphics{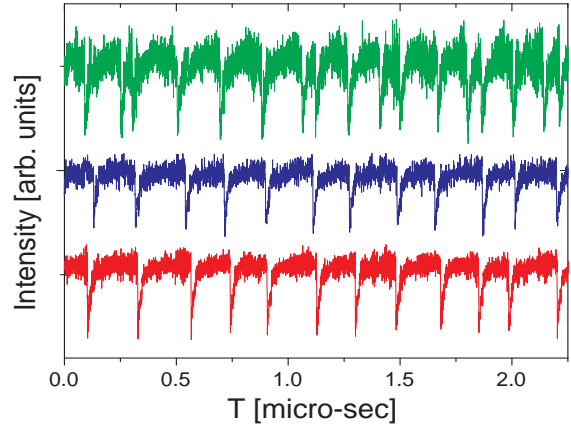}}}
\par}\vspace{-0.7cm}
\caption{\label{Laser2}A time sequence of all three laser
intensities. Lasers A and B are shown in blue and red while the
attacker, laser C, is in green. The attacker shows numerous failures
in its attempt to follow the well established correlation in LFF
breakdowns between the two parties.}
\end{figure}

A typical time sequence of the three laser intensities is shown in
Fig. \ref{Laser2}. The attacker, is shown in green while the
mutually coupled lasers, $A$ and $B$, are in red and blue. A
pronounced time correlation of LFF breakdowns is a  manifestation
of the synchronization of A and B. A detailed discussion of this
mutual synchronization will be given below. Now we concentrate on
the striking differences between the signal of laser $C$  and that
of lasers $A$ and $B$. Laser $C$ succeeds in synchronizing with A
and B occasionally, but frequently, additional breakdowns, which
are absent  in $A$ and $B$  appear in the sequence of $C$. Thus it
fails to follow the pair of mutually coupled lasers in a reliable
manner. The parameters of laser $C$, such as temperature, driving
current, feedback and coupling strengths were carefully adjusted
to be identical to the other two lasers. In practice we exercised
the following experimental procedure. We verified that laser $C$
can be synchronized unidirectionally to a single master (either
laser A or B). A fine tuning of all laser C parameters was made to
achieve the best synchronization possible . Then we allowed the
weak mutual coupling of lasers A and B and observed that the
synchronization of laser $C$ degrades immediately. Trying to fine
tune laser C parameters, so as to obtain better attacker
synchronization did not improve the quality of synchronization. We
were able to realize a high level of synchronization if either the
coupling of laser $C$ to the communicating pair was significantly
increased or  the feedback to laser C was considerably reduced.
But in this situation the C laser parameters were  so different
from the parameters of A and B that it could not possibly  be used
to decipher information as discussed below.

\begin{figure}
\vspace{-0.8cm}
{\centering
\resizebox*{0.5\textwidth}{0.3\textheight}
{{\includegraphics{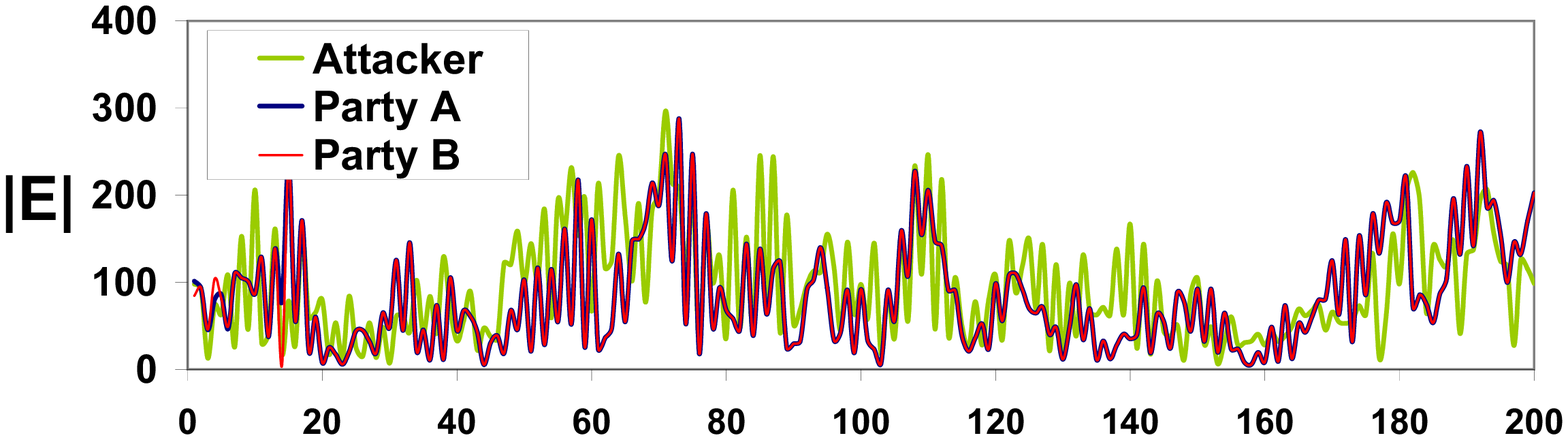}}}
\par}\vspace{-3.8cm}
\caption{\label{sim}Numeric simulation results. A typical sequence
of the amplitude $|E|$, without LFF, for the attacker and the
parties with $\kappa = \sigma = 10^{10}s^{-1}$. The signals of the
paries are overlapping and therefore are indistinguishable.}
\end{figure}

\begin{figure}
\vspace{-0.8cm}
{\centering
\resizebox*{0.5\textwidth}{0.3\textheight}
{{\includegraphics{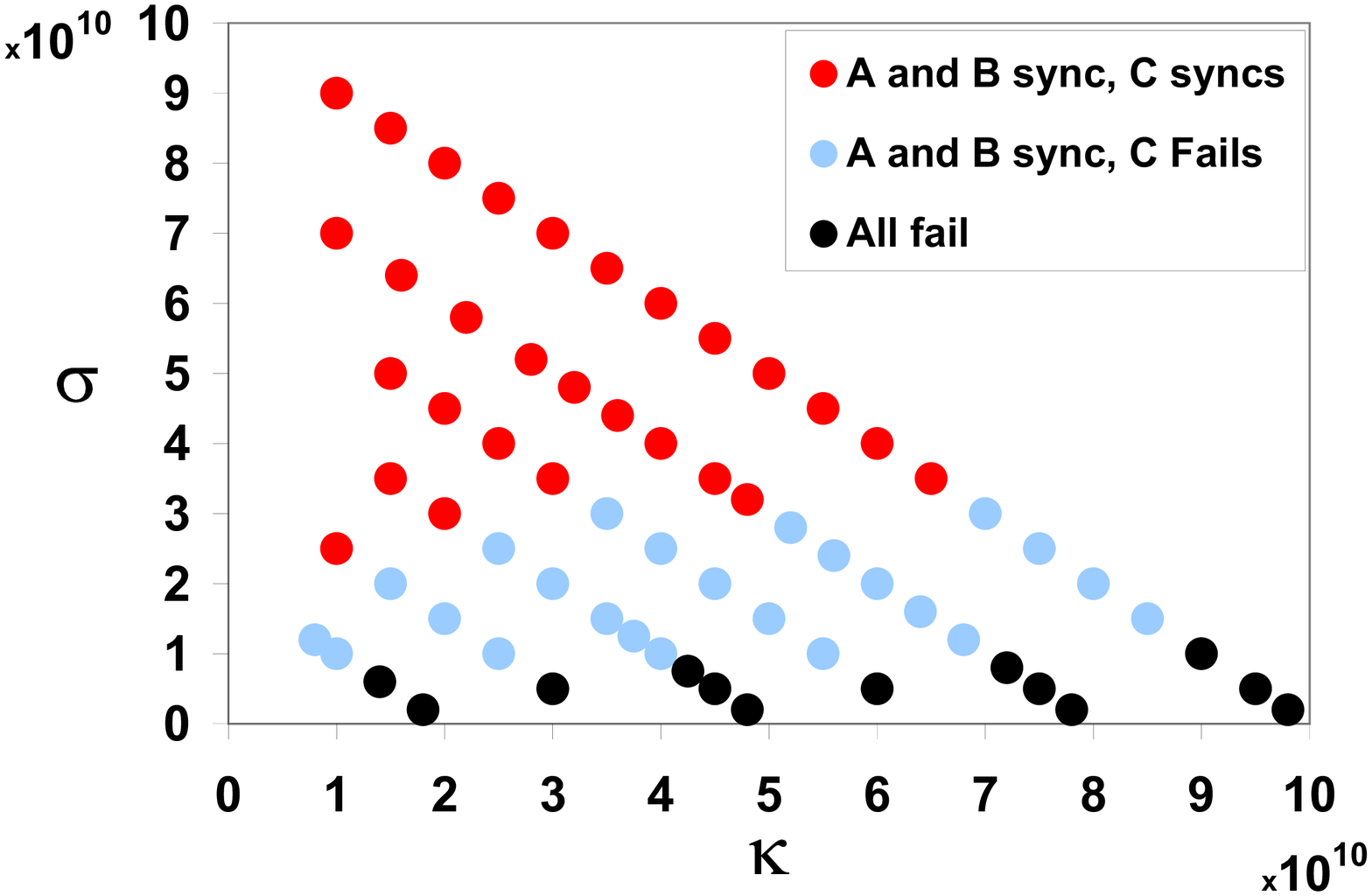}}}
\par}
\vspace{-1.cm}
\caption{\label{sim2}Success or failure of
synchronization for the parties and the attacker for a range of
parameter values $\kappa$ and  $\sigma$.}
\end{figure}

\begin{figure}
\vspace{-0.5cm} {\centering
\resizebox*{0.5\textwidth}{0.3\textheight}
{{\includegraphics{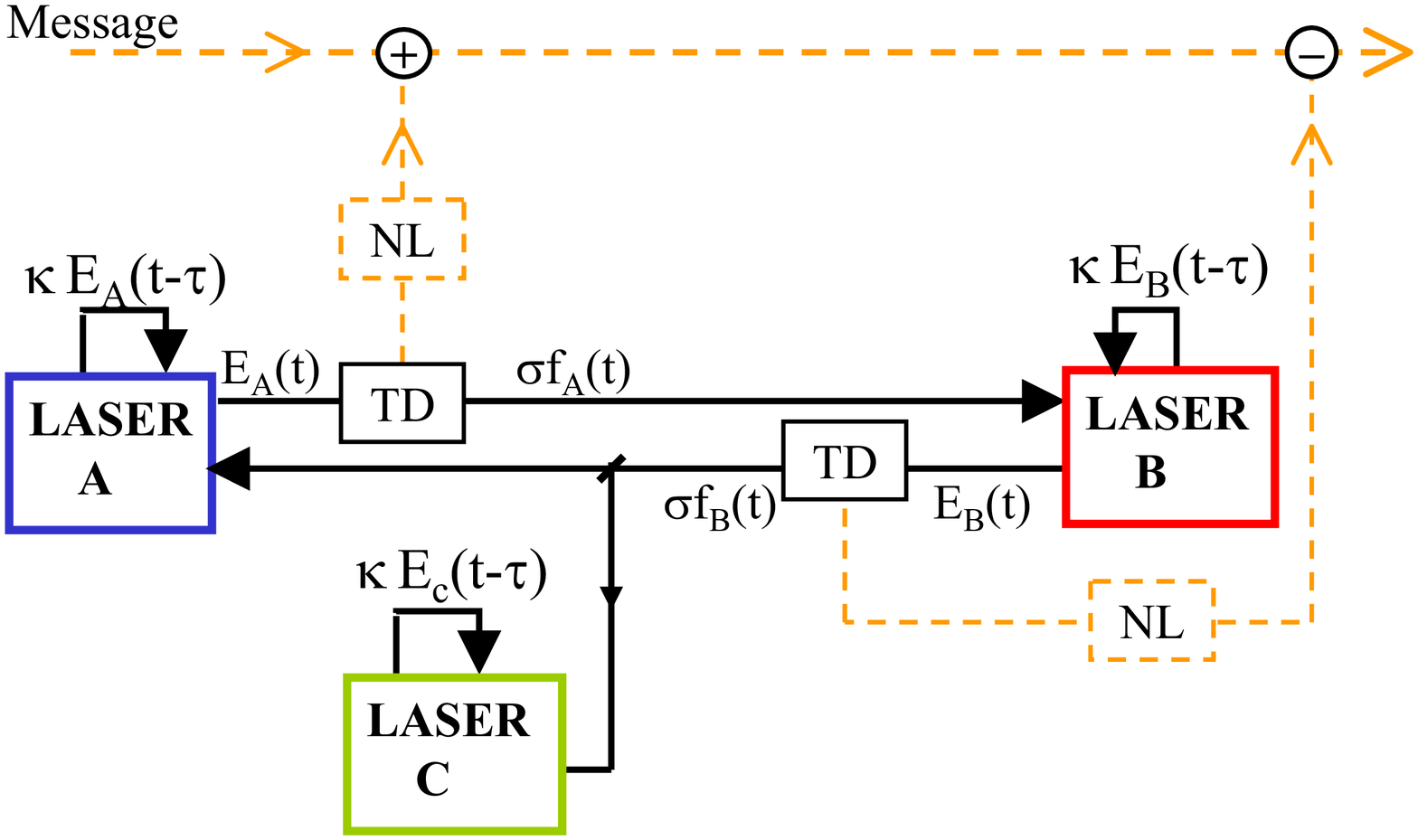}}}
\par}\vspace{-1.5cm}
\caption{\label{Simsetup}Schematic cryptographic setup . Parties
$A$ and $B$ and the attacker, $C$, receive external feedback of
strength $\kappa$. The mutual signals, $\sigma f_A(t)$ and $\sigma
f_B(t)$, consists of a linear combination of three delayed
signals. $\sigma f_B(t)$ is also used as an external input to the
attacker $C$. The naive way to cipher the message, using a linear
combination of time delayed signals of laser $A$ and nonlinearity
is depicted in orange.} \end{figure}

To model the single-mode semiconductor lasers we used the Lang
Kobayashi differential equations, as defined in \cite{alhers98}. For
the dynamics parameters we used the values in \cite{alhers98}. The
external feedback strength is defined by $\kappa$ and the coupling
strength of lasers A and B and the attacker to A or B is defined by
$\sigma$. The calculated sequences behaved exactly as those observed
experimentally in the LFF regime. The interval between breakdowns
can be controlled by $\kappa$ and $\sigma$. The synchronization of
lasers A and B is robust to small differences in the parameters
\cite{NoiseRobustness}. Furthermore lasers A and B were found to be
synchronized in between the LFF breakdowns. For small $\kappa$ and
$\sigma$ chaotic synchronization is also possible in the regime
without LFF breakdowns. Fig. \ref{sim} displays a typical sequence
of amplitudes of all three lasers in this regime with
$\tau_{c}=\tau_{d}=10ns$. Note that lasers A and B are
indistinguishable in the picture.


We now discuss the two dimensional phase space, defined by
parameters $\kappa$ and $\sigma$. The examined phase space is
characterized by the following three regimes as depicted in Fig.
\ref{sim2}. The red regime where $\sigma$ is strong enough in
comparison to $\kappa$, and all lasers are synchronized. The black
regime where the coupling is negligible and there is a lack of
synchronization between any of the lasers. Most interesting is the
window of the light blue regime where A and B are synchronized, but
C fails to synchronize. This regime represents an intermediate
window where mutual coupling is superior to a unidirectional
coupling \cite{LEremark}. This new effect is at the center of our
cryptographic system presented below.

In our cryptographic system, the coupling signal, $f_A(t)$, is a
linear combination of time delayed signals of the laser, $E_A(t)$,
as defined in Eq. 1 (and similarly for laser $B$). The initial
reason for using a complicated signal, $f_A(t)$, is that in this
way the original signal $E_A(t)$ remains secret and can be used to
mask a message. Using $f_A(t)$ as the coupling signal serves an
additional purpose: the attacker is forced to use {\it exactly}
the same coupling strength $\sigma$ as the parties. He cannot
amplify his coupling so as to force synchronization because when
he does so, he only synchronizes with $f_A(t)$ as the output of
the laser, instead of $E_A(t)$. Laser $C$ must have the same
``heart beats'' and the same internal ``flows'' in all of its
internal optical ''wires'', as the synchronized lasers, $A$ and
$B$, otherwise the beam it constructs to decrypt the message is
essentially different from $E_A(t)$. The use of  a combination of
time-delayed values as the coupling signal between the parties
thus accomplishes two necessary tasks: it hides the original
signal of the laser so that we can use it to mask a message, and
it prevents the attacker from using a stronger coupling so as to
synchronize.

A prototypical cryptographic setup  is shown in Fig.
\ref{Simsetup}. Each of the three lasers has external feedback of
strength $\kappa$ with time delay $\tau$. The transmitted signal
of each of the two lasers, $f_A(t)$ and $f_B(t)$, consists of a
linear combination of the time delayed
 laser signals

\begin{equation}
f_A(t)=E_A(t-\tau_1)+E_A(t-\tau_2)+E_A(t-\tau_3) ...
\end{equation}

\noindent and similarly for laser  $B$. In our simulations we have
seen that using a combination of time delays and nonlinearity does
not disturb the synchronization of A and B \cite{NoiseRobustness,
Losy}. The attacker, $C$, tries to imitate one of the mutually
coupled lasers, hence it will use as input $\sigma f_B(t)$, for
instance, and feedback characterized by $\kappa$ and $\tau$.

The part of the publicly known setup required to conceal the
message, in addition to the synchronization process is marked in
orange in Fig. \ref{Simsetup}. Two (or in general more than two)
delayed signals are summed and can be further complicated by
optional nonlinear terms. The resulting beam is used to cipher the
message, which is subsequently transmitted to the other party.
Party $B$, who is synchronized to party $A$, can easily generate
an identical beam to the one used to cipher the message.
Subtracting this beam from the received signal immediately reveals
the transmitted message.

A possible attack on the communication channel is via the use of a
chaos pass filter on the encrypted signal (the orange dashed line
in Fig. \ref{Simsetup}). In this strategy, the attacker uses the
encrypted transmitted message as an external input to a laser
whose output is expected to be the chaotic signal without residues
of the message \cite{CPF}. The attacker can then subtract this
chaotic signal from the encrypted message and reveal the message.
This strategy was shown to work with a simple, single frequency
continuous-wave message only \cite{elsasser01,CPF}. It would fail
if a broadband message is used such as a compressed message or
alternatively a message which is modulated over a large bandwidth.
In addition, if the message is ciphered with a non-linear function
of time delayed signals, and not with the simple laser output,
$E_A(t)$, it also prevents the successful application of chaos
pass filtering.

Another possible attack strategy is to derive the original signal
$E_A(t)$ from the transmitted signal $f_A(t)$, by using some
mathematical analysis method, such as embedding of the
time-delayed values. Such an embedding method was shown to be
unsuccessful, especially when stochastic ingredients were added to
the dynamics \cite{Losy}. Here too the lasers are governed by a
stochastic process, and thus a mathematical analysis is useless.

\begin{figure}
\vspace{-0.3cm} {\centering
\resizebox*{0.43\textwidth}{0.25\textheight}
{{\includegraphics{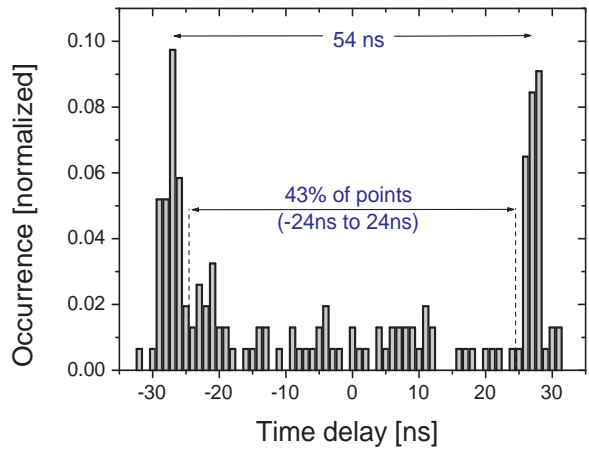}}}
\par}
\caption{\label{Laser3}Histogram of time delay/anticipation of power
breakdowns of two parties (lasers A and B). Two strong peaks at
exactly the coupling time between the two parties ($\tau_{c}$=27 ns)
include 57$\%$ of registered events (master/slave configuration).
However another 43$\%$ of the events fall between the peaks showing
synchronization regime beyond the master/slave configuration.}
\end{figure}

Let us now discuss the synchronization of two mutually coupled
lasers  in detail. A close examination of Fig. 2 (blue and red
traces) reveals that the LFF breakdowns are not perfectly
correlated up to the time window of the coupling lengths between
the lasers. This intriguing dynamics is summarized in an histogram
of power breakdown correlations as shown in Fig. 6. Two different
synchronization regimes can be readily distinguished. The first
one is characterized by a constant lag/lead time between power
breakdowns which corresponds to the coupling time $\tau_{c}$. In
Fig. 6 this corresponds  to the two narrow peaks ($\pm$27 ns)
which include about 57$\%$ of all registered events. This type of
synchronization known as a master/slave (leader/laggard)
configuration was shown in unidirectional coupling \cite{shore99}
as well as in mutual coupling without external
feedback\cite{elsasser01}. The heights of the two peaks are nearly
equal which means that the leading role is equally distributed
between the lasers (the system is symmetric). The second
synchronization regime, which covers another 43$\%$ of events is
characterized by a lead/lag time which is less than $\tau_{c}$ and
equally distributed in the time interval between 0 and $\pm
\tau_{c}$. In this situation neither of the lasers is master or
slave and the synchronization enters the regime which is beyond
the master/slave treatment. The mutually coupled lasers jump
randomly throughout both synchronization regimes. For the
communication scheme presented here the mutually coupled lasers
have to operate outside the leader/laggard regime. For this the
careful mode matching of two lasers achieved by fine tuning of
temperature and current is crucial. In our simulations we observed
exactly the same synchronization regimes. We saw that when using
$\tau_{c}= \tau_{d}$ the parties display only the second
synchronization (with no leader/laggard) \cite{NoiseRobustness}.

To conclude, we have presented a framework for a public-channel
cryptographic system, based on two mutually coupled lasers. The
system proposed here opens a manifold of possibilities. For
instance,  the extension of our framework to generate secret
communication among a group of more than two lasers. There are
many details in this systems that deserve further experimental and
theoretical research - the sensitivity of the synchronization to
the parameters such as time delays, non-linearity, the carefully
matched laser parameters  and the distance between the lasers.
Also the ciphering of the message using the synchronized signal,
deserves further investigation and is open to different
implementations.

The research is partially supported by the Israel Science
Foundation.


\begin{thebibliography}{9}
\bibitem{LEremark}In our simulations with two coupled Lorenz
systems \cite{Losy}, we have measured the conditional Lyapunov
exponent (CLE), and noticed that there is an area for which the CLE
is negative for the parties and positive for the attacker. This
means that even if the attacker manages to get close to the parties
at some point, he will after a short time become very far.
\end{thebibliography}
\end{document}